\documentclass[a4paper]{rsauthor}
\usepackage[a4paper]{geometry}
\usepackage[sort&compress,numbers]{natbib}
\usepackage{array}
\usepackage{amssymb}
\usepackage{amsmath}
\usepackage{color}
\usepackage{bm}
\usepackage{graphicx}

\newcommand{\ex}[1]{\left<#1\right>}
\newcommand{\cen}[1]{\tilde#1}
\newcommand{\floor}[1]{\lfloor#1 \rfloor}
\newcommand{\ceil}[1]{\lceil#1 \rceil}

\jname{Proc. R. Soc. A}

\title{The inefficiency of re-weighted sampling and the curse 
      of system size in high order path integration}
\author{Michele Ceriotti$^1$, Guy A. R. Brain$^1$, Oliver Riordan$^2$, and David E. Manolopoulos$^1$}
\address{$^1$Physical and Theoretical Chemistry Laboratory, 
South Parks Road, Oxford OX1 3QZ, UK 
$^2$Mathematical Institute, 24-29 St Giles', Oxford OX1 3LB, UK}
\date{\today}
\abstract{
   Computing averages over a target probability density by 
   statistical re-weighting of a set of samples with a different
   distribution is a strategy which is commonly adopted in 
   fields as diverse as atomistic simulation and finance. 
   Here we present a very general analysis of the accuracy and
   efficiency of this approach, highlighting some of its 
   weaknesses. We then give an example of how our results
   can be used, specifically to assess the feasibility of high-order 
   path integral methods. We demonstrate that the most promising
   of these techniques -- which is based on re-weighted sampling --
   is bound to fail as the size of the system is
   increased, because of the exponential growth of the
   statistical uncertainty in the re-weighted average.
}
\begin{document}   
   \maketitle
   \section{Introduction}
   
   Averaging is to the theoretical and computational sciences what
   measuring is to empirical science, and it is hard to imagine
   a problem with more general implications. Formally, the problem
   of computing an expectation value for a quantity $a(x)$ over a given 
   probability distribution $p(x)$ can be 
   stated as 
   \begin{equation}
      \left<a\right>_p = \int a(x) p(x) \mathrm{d} x.
      \label{eq:int-average}
   \end{equation}
   
   Very often, it turns out that $x$ spans 
   a very high dimensional space, which makes it impossible to 
   evaluate the integral on a grid of points. Instead, importance
   sampling algorithms \cite{metr+53jcp,hast70bio,frenk-smit02book}  
   are used which generate a set of points
   $x^{(p)}_i$ distributed in accordance with the target probability $p$. 
   Then, the expectation value can be computed as an average over 
   that set of points,
   \begin{equation}
      \left<a\right>_p \approx \bar{a}_n^{(p)}= 
      n^{-1}\sum_{i=1}^{n} a\left(x^{(p)}_i\right).
      \label{eq:is-average}
   \end{equation}
   
   The error in the estimate~(\ref{eq:is-average}) decreases 
   with $n^{-1/2}$, \emph{provided that the sample points 
   are uncorrelated}. In a number of circumstances
   obtaining uncorrelated samples from $p$ is 
   an exceedingly difficult problem, either because generating
   an individual point is computationally demanding, or because
   the sampling algorithm produces strongly correlated points,
   so that each one contributes very little to the reduction of the
   statistical error. 
   
   In these cases one is led to explore the possibility
   of performing re-weighted sampling, i.e., generating points
   according to a different distribution $p_0(x)$, 
   which is easier to sample efficiently, and then computing
   the average relative to $p$ using a correction weight 
   $w(x)=p(x)/p_0(x)$ involving the 
   ratio of the probability densities:
   \begin{equation}
      \left<a\right>_p=\frac{\left<a w\right>_{p_0}}{\left<w\right>_{p_0}} \approx 
      \bar{a}_n =
      \frac{\sum_{i=1}^{n} a\left(x^{(p_0)}_i\right) 
      w\left(x^{(p_0)}_i\right)}
      {\sum_{i=1}^{n} w\left(x^{(p_0)}_i\right)}.
      \label{eq:rw-average}
   \end{equation}
   Many methods have been developed which are fundamentally based on
   the re-weighted average~(\ref{eq:rw-average}), 
   which has found applications in physical chemistry \cite{duan+87plb,kuma+92jcc,bono+09jcc},
   theoretical physics \cite{barb+98npb,fodo-katz02plb}, economics and statistics \cite{hast70bio,smit-robe93jrssb}.
   Section~\ref{sec:re-weighting} will be devoted to a thorough analysis 
   of the statistical properties of re-weighted sampling, and 
   we will discuss under quite general assumptions the conditions required
   to apply this technique successfully.
   
   Our analysis allows one to assess the efficacy of re-weighting
   in each of the contexts in which it is applied. One instance
   of an important problem which has recently been tackled by
   re-weighted sampling concerns the use of 
   high-order path integration techniques in 
   computer simulations which allow for
   the quantum nature of atomic motion. 
   Conventional path integral (PI) methods \cite{feyn-hibb65book,parr-rahm84jcp,cepe95rmp} 
   involve a significant overhead compared to a purely classical treatment, and a 
   considerable effort has been devoted to the quest for less
   computationally demanding approaches. High order discretizations
   of the path have been available for some time \cite{taka-imad84jpsj,suzu95pla,chin97pla}, 
   but it has only recently become apparent that one can
   circumvent the calculation of  the bothersome second derivatives
   of the physical potential that arise in these discretizations by using
   re-weighted sampling \cite{jang+01jcp,yama05jcp}. 
   If successful, this approach would
   make high order path integration generally applicable and
   very attractive for many applications.
   
   To provide an example of the utility of the results established in Section~2, 
   we shall therefore    
   proceed in Section~3 to focus on the role of re-weighted 
   sampling in high order PI simulations. 
   We shall show in particular that this  form
   of path integration is eventually bound to fail because of the 
   reduced statistical efficiency introduced by the re-weighting. 
   Simulations of small clusters and/or mildly quantum
   mechanical problems constitute a niche in which this approach 
   might be beneficial \cite{yama05jcp},
   but its performance is bound to degrade for larger systems.

   \section{Statistics of re-weighting}
   \label{sec:re-weighting}
   
   A key issue which is often overlooked is that averaging
   according to~(\ref{eq:rw-average}) will have a lower statistical 
   efficiency than sampling $p$ directly as in~(\ref{eq:is-average}),
   i.e., for the same number of \emph{uncorrelated} sample points, 
   the statistical error in the re-weighted average will often be larger.
   Working out a simple, analytical expression for this drop in 
   sampling efficiency would be extremely useful, as it would 
   allow one to make an informed choice as to whether the 
   computational gain of sampling $p_0$ is 
   overshadowed by the concomitant statistical inefficiency.

   In order to evaluate such an estimate, let us first simplify the notation
   of Eq.~(\ref{eq:rw-average}) by labelling $a_i$ the $i$-th 
   sample for the observable $a$, and
   $w_i$ the ratio of the probability densities $p(x_i)/p_0(x_i)$.
   Averages with respect to the target distribution $p$ will be written as 
   $\left<\cdot\right>_p$, while averages relative to $p_0$ will be written
   as $\left<\cdot\right>$. With this notation, 
   \begin{equation}
      \bar{a}_n=\sum_{i=1}^n a_i w_i/\sum_{i=1}^n w_i,
      \quad \text{and}\quad \lim_{n\rightarrow\infty} \bar{a}_n=
      \frac{\ex{aw}}{\ex{w}}=\left<a\right>_p.
       \label{eq:an-short}
   \end{equation}
   Our objective is to unravel the statistical properties of the 
   weighted average~(\ref{eq:an-short})
   for a finite number $n$ of sample points, and to compare them 
   with those resulting from sampling the target distribution directly. 
   We will assume that different samples are completely uncorrelated,
   since different sampling distributions and algorithms
   can be compared on the basis of the computational cost of
   generating a new uncorrelated sample point.
   
   It is shown in the appendix that an asymptotic expression
   for the expectation value and the variance of $\bar{a}_n$ 
   can be obtained under very weak assumptions about    
   the joint probability distribution
   \[
      p_0(a,w)=\int \delta\left(a-a(x)\right)\delta\left(w-w(x)\right) p_0(x) \mathrm{d}x
   \]
   of $a$ and $w$. To first order in $n^{-1}$,
   \begin{eqnarray}\label{eq:sx-an1}
    \ex{\bar{a}_n} &\approx& \frac{\ex{aw}}{\ex{w}} + 
    \frac{\ex{aw}\ex{w^2}-\ex{w}\ex{aw^2}}{n\ex{w}^3}, \\
    \label{eq:sx-an2}
 \sigma^2(\bar{a}_n) &\approx&\frac{\ex{a^2w^2}\ex{w}^2-2\ex{aw}\ex{aw^2}\ex{w}+\ex{aw}^2\ex{w^2}}{n\ex{w}^4}.
   \end{eqnarray}
   Eq.~(\ref{eq:sx-an1}) shows that for any finite number of samples
   the re-weighted average is a biased estimator of $\ex{a}_p$, i.e., that
   $\left<\bar{a}_n\right>$ is affected by a systematic error which decreases with $n^{-1}$,
   while Eq.~(\ref{eq:sx-an2}) shows that the statistical error in the re-weighted mean 
   decreases asymptotically as $n^{-1/2}$.
   
   In order to perform a  comparison between the weighted 
   and the un-weighted sampling strategies, it is necessary to
   evaluate the coefficients which enter Eqs.~(\ref{eq:sx-an1}) and~(\ref{eq:sx-an2}).
   To do this we introduce 
   an additional assumption, namely that $a$ and $h=-\ln w$ are correlated 
   Gaussian variates, with means $\ex{a}$ and $0$, 
   variances $\sigma^2(a)$ and $\sigma^2(h),$ and covariance 
   $\left<a h\right>$, all with respect to $p_0$.  We introduce $h$ because in many cases -- 
   certainly the vast majority of those occurring in the physical 
   sciences -- the probability density is proportional
   to the exponential of an extensive state function
   (e.g. a difference Hamiltonian). 
   It is not difficult to justify the assumption of Gaussian statistics, since 
   the observable and the logarithm of the weight are often
   computed by summing a large number of fluctuating contributions, and 
   will therefore have nearly Gaussian statistics by virtue of the
   central limit theorem. In any case the objective here is simply to discuss
   the qualitative features of re-weighted sampling, and the Gaussian limit
   provides a convenient framework in which to do this.
   
   Under the assumption of a Gaussian joint probability distribution for $\left(a,h\right)$, 
   all expectation values of the form $\ex{a^p w^q}$ can be computed analytically in terms
   of the parameters of the distribution.  Eqs.~(\ref{eq:sx-an1}) and~(\ref{eq:sx-an2}) then 
   simplify to
   \begin{eqnarray}
      \left<\bar{a}_n\right> &\approx &\left<a\right>-\left<a h\right>+\left<a h\right> e^{\sigma^2(h)}/n,
      \label{eq:an-avg} \\
      \sigma^2(\bar{a}_n) &\approx&
      \left(\sigma^2(a) +\left<a h\right>^2\right) e^{\sigma^2(h)}/n.
      \label{eq:an-err}      
   \end{eqnarray}
   These equations convey in a more transparent way 
   the message that is inherent in their more general counterparts. 
   The systematic bias depends on the cross-correlation between the observable $a$
   and $h=-\ln w$, and both the systematic and statistical error grow
   exponentially with $\sigma^2(h)$.
      Whenever the fluctuations of $h$ are smaller
   than $1$, there will be little difference between the efficiency
   of sampling directly based on the target distribution and that of the
   re-weighted method. However, as soon as $\sigma(h)$ becomes
   larger than one, both the systematic bias and the statistical error
   of the re-weighted approach will 
   require a much larger number of samples for convergence to within 
   a given threshold.

   \begin{figure}[hbt]
      \centering\includegraphics{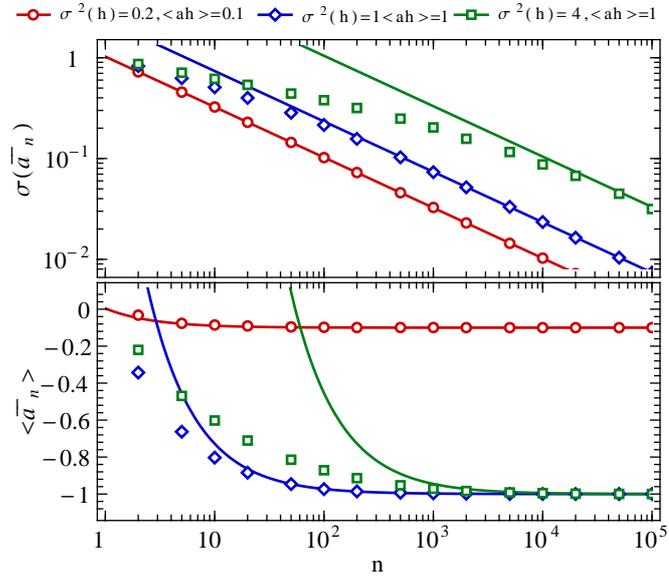}
      \caption{Numerical results for the mean and standard deviation in the
      mean for $n$  independent samples obtained from re-weighting a
      Gaussian-distributed observable with zero mean and variance
      $\sigma^2(a)=1$,
      based on a difference Hamiltonian $h$ with variance $\sigma^2(h)$, and
      cross-correlation $\left<a h\right>$ with $a$. The full lines correspond
      to the predictions of Eqs.~(\ref{eq:an-avg}) and~(\ref{eq:an-err}).    }
      \label{fig:gauss-tests}
   \end{figure}
   
     In Figure~\ref{fig:gauss-tests} we present the results of some numerical
   experiments in which we generated correlated Gaussian variables $a$ and $h$,
   and computed the weighted average of $a$ according to~(\ref{eq:rw-average}). 
   The results confirm the asymptotic nature of our estimates, which describe
   the behaviour of $\left<\bar{a}_n\right>$ and $\sigma^2(\bar{a}_n)$
   in the large-$n$ regime. The small-$n$ limit
   can instead be pin-pointed very easily by considering the $n=1$ case, in which
   the average converges to the un-weighted value $\left<a\right>$, and the variance
   is just $\sigma^2(a)$.

   While it is common knowledge that the fluctuations of the ``difference Hamiltonian''
   $h=-\ln w$ must be small in order to apply re-weighting successfully, 
   we believe it will come as a surprise to many that fluctuations of the order
   of a few units in $h$ will decrease the sampling efficiency by orders of 
   magnitude. 
   The message here is that in re-weighted simulations
   the joint probability distribution of $a$ and $h$ 
   should be monitored carefully. The value of $\exp[{\sigma^2(h)}]$ provides
   a direct indication of the statistical inefficiency
   due to re-weighting, and the cross-correlation 
   $\left<ah\right>$ provides an equally direct indication of the systematic bias in
   $\left<\bar{a}_n\right>$.
   Both of these are straightforward to converge, as they involve
   un-weighted averages.

   While Eqs.~(\ref{eq:an-avg}) and~(\ref{eq:an-err}) cannot
   be used quantitatively in the case of non-Gaussian statistics, 
   they provide useful guidelines to judge the quality of sampling
   in a re-weighted calculation. Whenever the fluctuations of $h$ are large
   and the weighted average is much closer to $\left<a\right>$ than to 
   $\left<a\right>-\left<ah\right>$, it is very likely
   that the numerical results obtained from~(\ref{eq:rw-average}) will be meaningless.
   In order to provide an illustration of the utility 
   of these estimates, we will now show that they can be used to predict 
   the situations in which high-order PI schemes 
   based on re-weighted sampling can be used successfully, and those in which they cannot.

   \section{High order path integration}
   \label{sec:sc-pimd}
   
   \subsection{Suzuki-Chin path integration}
   
      Path integral methods aim at approximating the quantum mechanical
   partition function, $Z=\mathrm{Tr} e^{-\beta H}$, by factoring it 
   in such a way that the resulting expression can be evaluated within a classical framework.
   The simplest such factorization makes use of the
   well-known Trotter product formula \cite{schu96book}, and results in an expression
   which is equivalent to a purely classical partition function in an extended phase space
   consisting of many replicas (beads) of the system, connected by
   harmonic springs to form a closed path (necklace) \cite{ parr-rahm84jcp,cepe95rmp,chan-woly81jcp,feyn-hibb65book} .
   The overall Trotter Hamiltonian reads
   \begin{equation}\label{eq:thamiltonian}
     H^{(T)} =
     \sum_{j=1}^P
     \frac{p_j^2}{2m} + \frac{m}{2}
     \left(\frac{\beta \hbar}{P}\right)^2
     (q_j- q_{j+1})^2 +V(q_j),
   \end{equation}
   where $\{q_j\}$ and $\{p_j\}$ are the coordinates and momenta
   of the different beads, and $V(q)$ is the physical potential.
   The classical partition function computed with this Hamiltonian at 
   $P$ times the physical temperature converges to the fully quantum
   mechanical partition function $Z$, with an error
   that is second-order in $\beta/P$. The number of
   beads necessary for convergence is in general a small multiple of
   $\beta\hbar\omega_{max}$, where $\omega_{max}$ is the fastest
   physical vibration in the system. The computational cost therefore becomes
   very large for low temperatures and/or stiff vibrational modes,
   and considerable effort has been devoted to the quest for faster-converging 
   factorizations in order to reduce this cost.
   
   It has been shown \cite{chin06pla} that
   factorizations with an error that is of an order higher than two
   in $\beta/P$ must contain composite operators which have the form
   of nested commutators between the kinetic energy operator $\hat{T}$ and
   the potential $\hat{V}$. The complexity of the nested commutators
   increases with the order of the factorization and is only really
   manageable up to fourth-order.  Among the different fourth-order
   factorizations developed so far \cite{taka-imad84jpsj,li-brou87jcp,suzu95pla,chin97pla,jang+01jcp}, 
   the one proposed
   by Suzuki \cite{suzu95pla} and simplified by Chin \cite{chin97pla} 
   has better convergence properties than most others. It also only
   contains the double commutator $[\hat{V},[\hat{T}, \hat{V}]]$, which is proportional
   to the square modulus of the force and hence of a manageable complexity.
  
   This Suzuki-Chin (SC) factorization results in a Hamiltonian
   \begin{equation}\label{eq:schamiltonian}
     H^{(SC)} =
     \sum_{j=1}^P
     \frac{p_j^2}{2m} + \frac{m}{2}
     \left(\frac{\beta \hbar}{P}\right)^2
     (q_j- q_{j+1})^2 +\tilde{V}_j(q_j,\beta)
   \end{equation}
   which contains a bead-dependent modified potential
   \begin{equation}\label{eq:scpotential}
     \tilde{V}_j(q_j, \beta) = w_jV(q_j) +
     \frac{w_j d_j}{m}\left(\frac{\beta \hbar}{P}\right)^2
     \left|V'(q_j)\right|^2.
   \end{equation}
   The coefficients $w_j$ and $d_j$ take different values
   for odd and even beads, namely $w_{2j}=2/3$, $w_{2j+1}=4/3$,
   $d_{2j}=\alpha/6$, $d_{2j+1}=(1-\alpha)/12$, where $\alpha$ is
   a free parameter which can be varied in the range $[0,1]$.
   While tuning $\alpha$ can improve the convergence in specific
   cases \cite{jang+01jcp,yama05jcp}, there is no universally optimal choice, so in the present
   work we used $\alpha=0$.

   The modified Hamiltonian resulting from the SC
   factorization contains the square modulus of the physical force.
   A path integral molecular dynamics (PIMD) calculation based on the corresponding    
   classical equations of motion therefore requires one to evaluate
   the derivative of the squared force term, which contains the Hessian. 
   Unfortunately, this is prohibitively expensive to evaluate for large systems 
   with all but the simplest inter-atomic potentials.
  One way around this difficulty is to exploit the possibility
   of sampling based on a quasi-Trotter Hamiltonian which resembles
   Eq.~(\ref{eq:thamiltonian}) but has the physical potential on each bead
   weighted by $w_j$ \cite{jang+01jcp}.
   The statistics corresponding to the SC Hamiltonian can then be
   recovered by re-weighting configurations with the weighting factor
   $w=e^{-h}\equiv \exp(-\beta \Delta H/P)$, 
   where the difference Hamiltonian has the form
   \begin{equation}\label{eq:sc-deltah}
     \Delta H = \sum_{j=1}^P
     \frac{w_j d_j}{m} \left(\frac{\beta \hbar}{P}\right)^2\left|V'(q_j)\right|^2.
   \end{equation}
   The effect that re-weighting configurations based on~(\ref{eq:sc-deltah})
   has on the statistical efficiency of sampling is the fundamental
   drawback of this approach, as will be discussed in the next subsection,
   based on the general results of Section~\ref{sec:re-weighting}.
   
   \subsection{The curse of system size}
   \label{sec:size-curse}
   
   The high-order path integral strategy we have just described may appear to be 
   superior to Trotter PI in all respects \cite{jang+01jcp}.
   However, one must not overlook the question of statistical efficiency,
   which could imply that longer trajectories are required to obtain the 
   same level of sampling accuracy, a factor which must be considered when
   comparing the computational viability of different approaches.
   
   Given the analysis in Section~\ref{sec:re-weighting}, 
   all we need to evaluate the errors due to finite sampling
   in this context is an estimate of the variance of 
   $\Delta H$. Consider first the case of a simple harmonic oscillator with 
   frequency $\omega$, in the strongly-quantum regime, where the distribution of
   configurations is well approximated by the ground-state density
   $P(q)\propto \exp(-m\omega q^2/\hbar)$. Then, the squared fluctuations of
   $|V'(q)|^2$ read 
   \[
     \left<|V'(q)|^4\right>-\left<|V'(q)|^2\right>^2=\hbar^2 m^2 \omega^6/2.
   \]
   Under the simplifying assumption that the statistics of $q$ for 
   a finite-$P$ path integral simulation are 
   the same as those of the quantum oscillator at full convergence, one
   can write down the squared fluctuations of the 
   temperature-scaled difference Hamiltonian $h\equiv \beta \Delta H/P$ as
   \begin{equation} \label{eq:fluct-harm}
     \sigma^2(\beta \Delta H/P) 
     \approx (P/2)^k(\beta\hbar \omega/P)^6/162,
   \end{equation}
   where $k$ varies between one (if one assumes that there is no 
   correlation between the forces on different beads) and two (if
   one assumes that the forces on all beads are the same).  
   
   Eq.~(\ref{eq:fluct-harm}) implies that it is possible to reduce 
    the fluctuations of the weights $w=e^{-h}$ at will
   by increasing the number of replicas. However, one should remember that
   keeping this number low was the primary reason for attempting
   a high-order factorization in the first place.
   Since the relative error in the energy due to the discretization of the 
   fourth order PI grows as $(\beta\hbar \omega/P)^4$, it is clear that -- for a given 
   discretization error -- the sampling problems inherent in 
   re-weighted SC PIMD are bound to get worse as the quantum 
   nature of the problem becomes more pronounced.
   
   Moreover, the issue of statistical efficiency is exacerbated once one
   considers a system with $f$ degrees of freedom. Since the difference
   Hamiltonian is size-extensive, its variance will scale linearly with
   $f$, leading to an exponential dependence of the statistical and
   systematic finite-sampling errors on the size of the system. 
   For a multi-dimensional harmonic system with frequencies $\omega_i$
   one can estimate -- in the worst-case scenario of complete 
   correlation of the forces on different replicas -- 
   that the total squared fluctuations will be 
   of the order of 
   \begin{equation} \label{eq:fluct-many}
     \sigma^2(\beta \Delta H/P)\approx
     \left(\frac{\beta\hbar \bar{\omega}}{P}\right)^6 \frac{f P^2}{648}, \quad
     \bar{\omega}=\left({\frac{1}{f}\sum_{i=1}^f \omega_i^6}\right)^{1/6}.
   \end{equation}
   
   In practice, this means that the high-order path integral integration 
   scheme which we have described in the previous section
   will yield a small statistical 
   uncertainty up to a critical system size, above which the statistical
   efficiency will decrease exponentially.
   Our analysis of the harmonic limit yields an estimate of this critical size
   for a given value of $\beta\hbar\bar{\omega}$. 
   For instance, flexible water 
   can be simulated successfully by Trotter PI with $P\gtrsim 32$, and has   
   $ \bar{\omega}\approx 2700$~cm$^{-1}$. For $T=300$~K and 
   $P=32$, Eq.~(\ref{eq:fluct-many}) suggests that the fluctuations in 
   the difference Hamiltonian will be of the order of one in a simulation 
   of a few tens of water molecules. This is the sort of system size up to 
   which re-weighted SC PIMD is therefore predicted to be advantageous.
      
      \begin{figure}[hbt]
      \centering\includegraphics{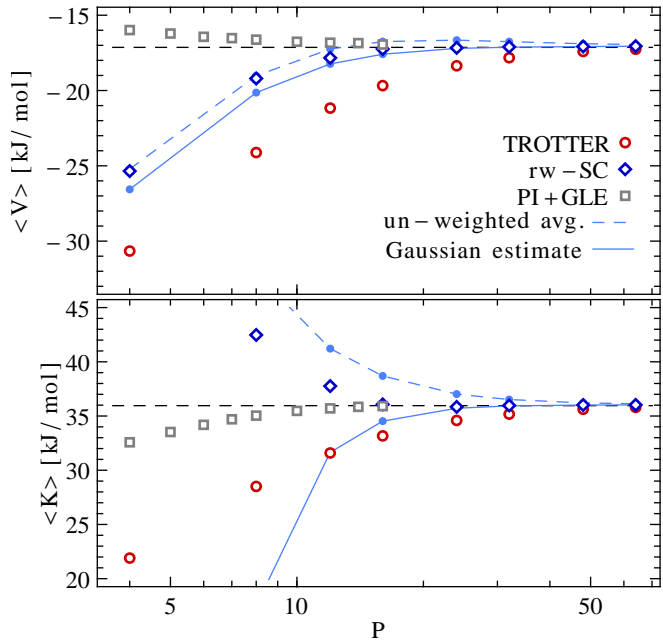}
      \caption{The two panels show the convergence of the 
      average potential and kinetic energy for a simulation of 
         $216$ water molecules as a function of the number 
         of beads $P$. The reference value -- shown as a black dashed line
      -- corresponds to Trotter PI with $P=128$. The results
      from the PI-GLE accelerated Trotter PIMD are also reported \cite{ceri+11jcp}.
      Error bars were calculated, but are smaller than the data points on this scale. 
      For the SC simulations we also report the 
      averages of the observables computed without weights (joined by dashed 
      lines) and those obtained from the infinite $n$ limit of  
      Eq.~(\ref{eq:an-avg}) (continuous lines).  }
      \label{fig:water-pconv}
   \end{figure}

   \begin{figure*}[hbt]
      \centering\includegraphics[width=1.0\textwidth]{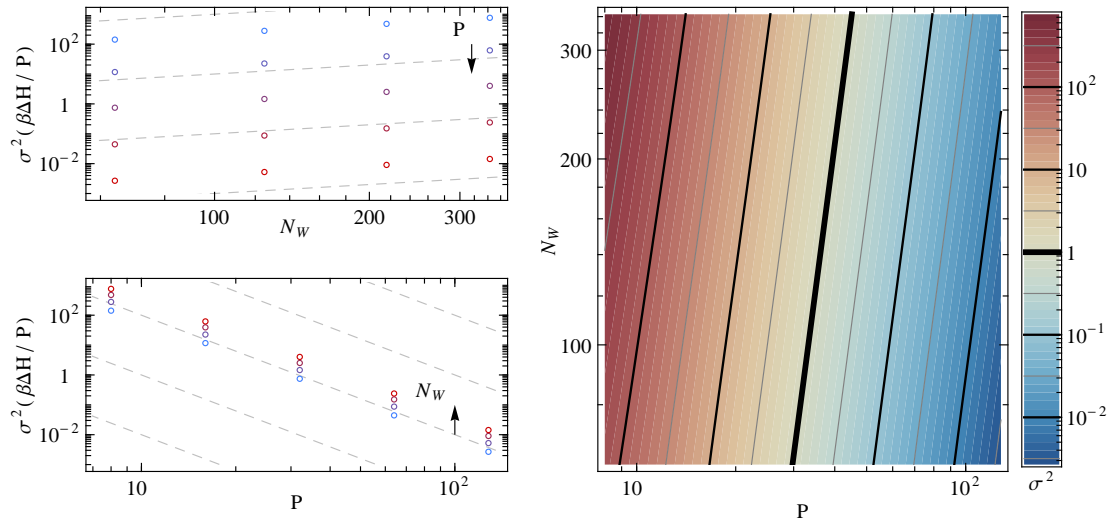}
      \caption{The three panels show the fluctuations of the
      temperature-scaled difference Hamiltonian $h=\beta\Delta H/P$ 
      for a simulation of a flexible water model at $T=298$~K, 
      for different numbers of replicas $P$ and numbers of
      water molecules $N_W$. The top left panel shows the data points
      as a function of $N_W$; the dashed lines are guides
      for the eye corresponding to $\sigma^2 \propto N_W$ 
      behaviour. The bottom left panel shows the points as a function of 
      $P$; here the dashed lines correspond to $\sigma^2 \propto P^{-4}$. 
      The right panel shows a contour plot of the interpolant
      of the data points as a function of $P$ and $N_W$, with 
      $\sigma^2$ denoted by the colour scale. 
      Red indicates regions in which the statistical
      inefficiency of re-weighted Suzuki-Chin PIMD is unmanageable, 
      while blue indicates regions in which it is comparable to 
      that of Trotter PIMD. Note that for a given $P$, the inefficiency
      gets worse with increasing system size.
        }
      \label{fig:water-dh2}
   \end{figure*}
   
   \begin{figure}[hbt]
   \centering\includegraphics{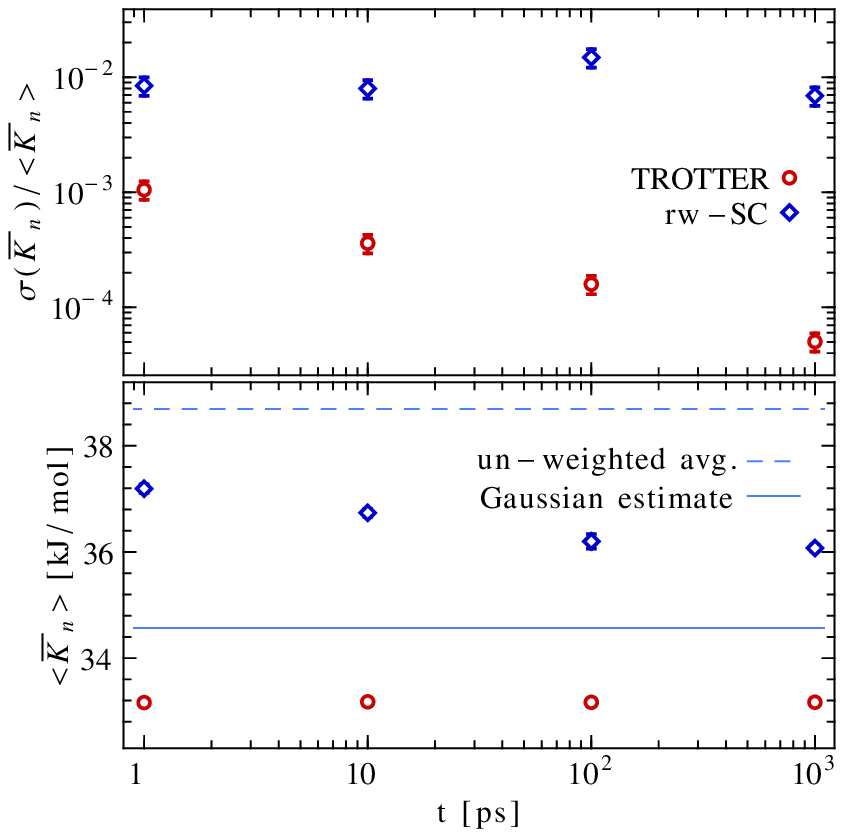}
   \caption{The two panels show the convergence of the 
   mean kinetic energy for a simulation of 
   $216$ water molecules, at $T=298$~K and for $P=16$, 
   as a function of the length of the 
   simulation, together with the relative fluctuations in the mean. 
   In the lower plot, the un-weighted average kinetic energy and 
   that obtained from the infinite $n$ limit of  Eq.~(\ref{eq:an-avg}) are also shown.}
   \label{fig:water-nconv}
   \end{figure}
   
   \subsection{An example: liquid water}
   \label{sec:liquid-water}
   
   In order to verify the predictions of Eq.~(\ref{eq:fluct-many})
   and those of Section~\ref{sec:re-weighting} for a realistic 
   condensed-phase example, we have examined the case of a flexible
   water model \cite{habe+09jcp}, simulated at a temperature of 298~K and 
   at the experimental density. We chose to perform simulations 
   with $216$ water molecules, so as to be in a regime in which -- 
   according to the arguments of Section~3b --
   the statistical inefficiency of re-weighting should be apparent.
   
   We explored numbers of beads ranging between $P=4$ and $P=128$, and
   for each set of parameters performed $16$ independent simulations,
   each of which comprised $20$~ps for the initial equilibration and 
   $1$~ns during which averages were accumulated.
   The calculations were accelerated using the ring-polymer contraction 
   technique \cite{mark-mano08cpl}, with the same parameters as we have employed in 
   previous simulations \cite{ceri+11jcp}. Canonical sampling was
   enforced by applying targeted Langevin thermostats to the internal
   modes of the ring polymer, and a global stochastic velocity rescaling 
   with a time constant of 10~fs to the centroid \cite{ceri+10jcp,buss+07jcp}. 
  The quantum mechanical kinetic energy of the water molecules was accumulated using the 
  modified form of the centroid virial estimator proposed by Yamamoto \cite{yama05jcp}. 
  This estimator, which is based on a scaling of fluctuation 
  coordinates \cite{pred+03jcp},
  can be evaluated by finite differences thereby avoiding the need for high-order 
   derivatives of the potential.
   
   As an initial test, we examined the convergence of the average kinetic and potential
   energies of the liquid as a function of 
   the number of replicas, for both Trotter PI and re-weighted SC PI
   (see Figure~\ref{fig:water-pconv}). As expected, and as witnessed in 
   previous calculations with a similar water model \cite{jang+01jcp}, the higher
   order of the PI factorization does translate into faster convergence.
   However, for the simulations with $P\lesssim32$, 
   the statistical error in the mean for the re-weighted SC method was found to be 
   considerably larger than that of Trotter PIMD. 
   Even more worrying, for re-weighted SC PIMD 
   there is a significant difference between
   the weighted average of the kinetic energy and the asymptotic result predicted
   for Gaussian statistics [the $n\to\infty$ limit of Eq.~(\ref{eq:an-avg})]. 
   These issues will be considered in more detail below.
   As a final remark, it can be seen from Figure~\ref{fig:water-pconv} 
   that the recently-developed stochastic  
   PI+GLE \cite{ceri+11jcp} approach manages to obtain even faster
   convergence than SC PIMD,  without any of the 
   sampling issues that we are about to elucidate.
   
   It remains to be verified whether the scenario presented in
   Section~3b based on a harmonic model and an
   analysis of the sampling efficiency in the Gaussian limit corresponds
   to the numerical finding in this real example. First of all, one
   needs to verify whether the scaling of the variance of the 
   logarithm of the weights follows Eq.~(\ref{eq:fluct-many}). 
   Figure~\ref{fig:water-dh2} demonstrates that both
   the scaling with $P$ and with the number of degrees of freedom
   correspond to the predictions we have made, even if the 
   quantitative values for $\sigma^2(\beta\Delta H/P)$ 
   are lower by about a factor of $3$ than the analytical estimates obtained 
   from the harmonic model. 
   Taking this correction into account increases to 
   about $100$ the number of water molecules for which
   re-weighted SC PIMD is computationally reasonable. 
   
   Having verified that the fluctuations in the difference Hamiltonian
   closely obey Eq.~(\ref{eq:fluct-many}),  
   let us discuss the validity of the relations between 
   these fluctuations and the error in the mean that we have 
   worked out in Section~\ref{sec:re-weighting}. 
   Figure~\ref{fig:water-nconv}
   shows that -- for the case of $P=16$, where the variance of 
   the temperature-scaled difference Hamiltonian $h=\beta\Delta H/P$ 
   is much larger than one -- the trends do indeed correspond
   to those derived in the limit of Gaussian statistics.
   Averages computed from short trajectories suffer a large systematic bias, 
   and they slowly approach the limit predicted from the cross-correlation
   between the observable and the difference Hamiltonian as the number 
   of samples is increased. To obtain a converged expectation value and therefore
   verify the accuracy of the prediction based on Gaussian statistics, 
   a prohibitively long trajectory would be required.
   Also as anticipated, the upper panel of Figure~\ref{fig:water-nconv}
   clearly demonstrates that the statistical variance in the mean for re-weighted
   SC PIMD decreases much more slowly than for Trotter PIMD, 
   completely offsetting the benefits of using the higher-order propagator.
   
   It is important to remark that we have been able to make these issues
   stand out clearly because we have considered a relatively simple 
   problem, for which we could run long trajectories and gather extensive
   statistics. In real applications one can rarely afford to 
   perform such a detailed error analysis, and in this respect the 
   analytical results we have derived provide a useful litmus test.
   The fluctuations of the temperature-scaled
   difference Hamiltonian $h$ should be smaller than one, since
   for larger fluctuations the statistical inefficiency rapidly
   becomes unmanageable.
   Even more interesting is the cross-correlation between the
   observable $a$ and $h$, which indicates the extent of the systematic bias.
   
   \section{Conclusions}
   
   In this paper we have performed a careful assessment of the 
   statistical efficiency of re-weighted sampling.
   Our analysis shows a dramatic degradation in 
   performance when the squared fluctuations of the logarithm
   of the weight exceed one. While it is common knowledge 
   that large fluctuations in the weights lead to failure
   of such techniques, and that the size-extensivity of the difference
   Hamiltonian makes these methods unsuitable for the study of large-scale
   problems, our analysis shows clearly how dramatic this 
   effect can be, and also that the problem is quite insidious.

   In fact, insufficient sampling in a re-weighted calculation is very difficult to detect.
   It mainly manifests itself through the error in the mean 
   decreasing more slowly than the inverse square root of the number
   of samples even after a very large number of samples have been accumulated,
   indicating that the asymptotic regime has not yet been reached
   (see the upper panel of Fig.~4). 
   An even greater  concern is that finite sampling implies a bias
   in the estimate of the observable, which decreases very slowly
   and would be exceedingly hard to detect in a more complex -- or 
   computationally demanding -- problem (see the lower panel of Fig.~4).
   
   As a demonstration of how these results can be used, we 
   have examined the applicability of high order path integral factorizations
   to condensed-phase simulations.
   The most promising of these techniques relies on 
   re-weighting based on a difference Hamiltonian \cite{jang+01jcp,yama05jcp}.
   We have worked out the dependence of the fluctuations in this quantity
   on the properties of the system being studied, the number
   of degrees of freedom involved, and the number of path integral beads. 
   We have shown that for large and/or strongly quantum systems,
   a large number of replicas is required, not so much in order to converge
   the asymptotic expectation values of quantum observables, but to keep the 
   fluctuations in the re-weighted average under control. 
   Our analytical results allow one to predict
   the break-even point at which the advantages of 
   high-order path integrals are lost due to statistical inefficiency.
   These predictions have been qualitatively verified by numerical experiments
   on the simulation of a flexible model of water under ambient conditions.
   For this system, high-order SC PIMD is advantageous for simulations of  
   up to around $100$ water molecules, but no more.
   
   While a niche for the use of Hessian-free high-order PIMD therefore exists for the 
   study of small clusters, one must consider that accelerated PIMD based on stochastic
   dynamics \cite{ceri+11jcp} exhibits comparable or faster convergence of quantum
   observables, without being affected by any of the statistical issues we have 
   focussed on here. We firmly believe that this stochastic acceleration is therefore 
   a far more promising approach to large scale path integral simulations.
      
\ack{
   The authors gratefully acknowledge an insightful discussion with 
   Michele Parrinello, and funding from the Wolfson Foundation, the 
   Royal Society, and the Swiss National Science Foundation.}
   
   \appendix
   
   \section{An asymptotic expression for re-weighted averages}
   \label{sec:formalproof}

In this appendix we present a derivation of the asymptotic
expansion of the expectation value of a re-weighted average of $n$
terms from a general distribution.
As usual, we say that the formal sum $\sum_{j=0}^\infty c_j n^{-j}$
is an asymptotic expansion for $f_n$ if for any fixed $k$ we have
$f_n=\sum_{j=0}^k c_j n^{-j} + \mathcal{O}(n^{-(k+1)})$ as $n\to\infty$.
This does not imply that the sum converges. Our main result is as follows:

\begin{theorem}
Let $(a_i,w_i)_{i=1}^n$ be $n$ independent samples from a (correlated) joint distribution
$p_0(a,w)$. Suppose that $w$ takes only positive values, and
that all moments of $a$, $w$ and $w^{-1}$ are finite.
Let
   \begin{equation}
      \bar{a}_n=\sum_{i=1}^n a_i w_i/\sum_{i=1}^n w_i.
   \end{equation}
Then $\ex{\bar{a}_n}$ and $\sigma^2(\bar{a}_n)=\ex{\bar{a}_n^2}-\ex{\bar{a}_n}^2$
have asymptotic expansions
in $n$ whose coefficients are determined by the joint moments of $p_0(a,w)$,
with the leading terms in these expansions being given by Eqs.~\eqref{eq:sx-an1} and~\eqref{eq:sx-an2}.
\end{theorem}

The argument below will show how to calculate the higher-order coefficients of the expansion, although there does not seem to be a simple explicit formula for the general 
coefficient $c_j$.

Multiplying the distribution $w$ by a constant (i.e., multiplying all
$w_i$ by the same constant) does not affect $\ex{\bar{a}_n}$, so we
assume in the proof that $\ex{w}=1$. With this assumption, Eqs.~\eqref{eq:sx-an1} and~\eqref{eq:sx-an2}
simplify to
\begin{eqnarray}\label{eq:an1}
 \ex{\bar{a}_n} &=& \ex{aw} +n^{-1}\left(\ex{aw}\ex{w^2}-\ex{aw^2}\right) + \mathcal{O}(n^{-2}), \\
\label{eq:an2}
 \sigma^2(\bar{a}_n) &=& n^{-1}\left(\ex{a^2w^2}-2\ex{aw}\ex{aw^2}+\ex{aw}^2\ex{w^2}\right)
 + \mathcal{O}(n^{-2}).
\end{eqnarray}
Explicit calculations can be further simplified by subtracting a 
constant from $a$ (and hence from $\ex{\bar{a}_n}$)
to ensure that $\ex{aw}=0$, but we shall not do this in the following.

   Since all the samples are independent and identically
   distributed, one can write
   \begin{equation}
      \left<\bar{a}_n\right>=
      n \left<\frac{a_nw_n}{z_n}\right>
 =       n \left<\frac{aw}{w+z_{n-1}}\right>
      \label{eq:compact-an}
   \end{equation}
   where $z_{n}=\sum_{i=1}^n w_i$ is the sum of $n$ independent 
   copies of $w$, and $z_{n-1}$ (but not $z_n$) is independent
of the correlated pair $(a,w)$.

Let $\cen{w}=w-1$ and $\cen{z}_n=z_n-n=\sum_{i=1}^n\cen{w}_i$
be the centralized versions of $w$ and $z_n$.
We first consider the properties of $\cen{z}_n$. 
Even in the (canonical) special case where $w$ is log-normal, not much
can be obtained analytically about the distribution of
$\cen{z}_n$ \cite{fent60iret}. However, one can obtain its moments using the general
formula
\begin{equation}
 \kappa^{(k)}=\mu^{(k)}-\sum_{i=1}^{k-1}\binom{k-1}{i-1}\kappa^{(i)} \mu^{(k-i)}
\label{eq:cum-mom}
\end{equation}
relating the cumulants $\kappa^{(k)}$ and moments $\mu^{(k)}$ of a distribution.
When $\mu^{(1)}=0$ (and hence $\kappa^{(1)}=0)$),
as is the case for the centred variables $\cen{w}$ and $\cen{z}_n$,
Eq.~\eqref{eq:cum-mom} can be written as
\begin{equation}
\cen{\kappa}^{(k)}  = \cen{\mu}^{(k)}- \sum_{i=2}^{k-2}\binom{k-1}{i-1}\cen{\kappa}^{(i)} \cen{\mu}^{(k-i)}.
\label{eq:cum-cmom}
\end{equation}
We therefore first calculate the cumulants $\cen{\kappa}_1^{(k)}$ of the distribution of 
$\cen{w}$ using Eq.~\eqref{eq:cum-cmom}. The cumulants of $\cen{z}_n$ are simply $\cen{\kappa}_n^{(k)}=n\cen{\kappa}_1^{(k)}$, and the moments $\cen{\mu}_n^{(k)}$ of $\cen{z}_n$ can be recovered by inverting Eq.~(\ref{eq:cum-cmom}). By induction on $k$ we find that $\cen{\mu}_n^{(k)}$ is a polynomial in $n$ of degree $\floor{k/2}$; this can also be seen directly by expanding the expectation of the $k$th power of the sum of the $\cen{w_i}$.

An asymptotic expansion for $\ex{\bar{a}_n}$ can now be developed by 
expanding the denominator in Eq.~(\ref{eq:compact-an}) about its mean.
Let $X=\cen{z_n}/n=z_n/n-1$. Using the identity $1/(1+x)=\sum_{j=0}^{k-1}(-x)^j+(-x)^k/(1+x)$, we have
   \begin{equation}
   \begin{split}
   \left<\bar{a}_n\right>=&
        \left<\frac{a_n w_n }{z_n/n}\right>=
        \left<\frac{a_n w_n }{1+X}\right>=\\
    =& 
      \sum_{j=0}^{k-1}(-1)^j\left<a_nw_n X^j\right>+
      (-1)^k\left<a_nw_n \frac{X^k}{1+X}\right>.
      \label{eq:trunc-series}
   \end{split}
   \end{equation}
Since $X=(\cen{z_{n-1}}+\cen{w_n})/n$ and $\cen{z}_{n-1}$ is independent of $(a_n,w_n)$,
the terms in the sum can be written out in terms of the 
moments $\cen{\mu}_{n-1}^{(k)}$ of $\cen{z}_{n-1}$:
   \begin{equation} 
      (-1)^j\left<a_nw_n X^j\right>=
      (-n)^{-j}\sum_{i=0}^j\binom{j}{i}\left<a(\cen{w}+1)\cen{w}^i\right>
      \cen{\mu}_{n-1}^{(j-i)}.
      \label{eq:awxj}
   \end{equation} 
For each $j$, the term 
described by Eq.~(\ref{eq:awxj}) is a (Laurent) polynomial in $n$
with powers between $n^{-\ceil{j/2}}$ and $n^{-j}$. 
We claim that 
\begin{equation}\label{eq:rnk}
 R_n^{(k)}=\left<a_n w_n \frac{X^k}{1+X}\right> = \mathcal{O}(n^{-k/2}),
\end{equation}
so the residual term $R_n^{(k)}$ shrinks with $n$. Moreover, 
summing Eq.~(\ref{eq:awxj}) over $j\le 2d$ and collecting coefficients
gives the first terms of the asymptotic expansion of $\ex{\bar{a}_n}$
down to order $n^{-d}$. For example, setting $d=1$ and discarding the 
terms with an $n^{-2}$ dependence yields Eq.~\eqref{eq:an1}. 

The bound Eq.~(\ref{eq:rnk}) is perhaps most easily seen by applying the Cauchy--Schwartz
inequality twice, or H\" older's inequality once, to show that in general
$\left|\ex{A_1A_2A_3A_4}\right|\le \prod_{i=1}^4 \ex{A_i^4}^{1/4}$.
Here $A_1=a_n$ and $A_2=w_n$ contribute constant factors to the product.
With $A_3=X^k=(\cen{z_n}/n)^k$ we have $\ex{A_3^4}=n^{-4k}\cen{\mu}_n^{(4k)}
=\mathcal{O}(n^{-2k})$, so $\ex{A_3^4}^{1/4}$ has the required order
$\mathcal{O}(n^{-k/2})$.
Finally, $A_4^4=(1+X)^{-4}=(z_n/n)^{-4}$.
Since the function $x^{-4}$ is convex, and $z_n/n$ is just the average
of the $w_i$, we have $A_4^4\le \sum_{i=1}^n w_i^{-4}/n$, so $\ex{A_4^4}\le \ex{w^{-4}}$,
which is by assumption a finite constant, and the bound Eq.~(\ref{eq:rnk}) follows.

For the variance, the argument is very similar, now using the fact that
\[
 \ex{\bar{a}_n^2} = n(n-1) \ex{\frac{a_{n-1}w_{n-1}a_nw_n}{n^2(1+X)^2}}+ n\ex{\frac{a_n^2w_n^2}{n^2(1+X)^2}},
\]
and the identity $(1+x)^{-2}=\sum_{j=0}^{k-1}(j+1)(-x)^j + (-1)^k\frac{(k+1)x^k+kx^{k+1}}{(1+x)^2}$. A similar method applies to higher moments of $\bar{a}_n$.

Note that the assumption that all negative moments of $w$ are finite was only used to control
the tail behaviour of $1/(1+X)=(z_n/n)^{-1}$ near 0. In fact, this has much better tail behaviour
than $w^{-1}$, so much weaker conditions suffice.
For example, splitting the sum $z_n$ into groups of $m$ and noting
that $z_m^{-1}\le \prod_{i=1}^m w_i^{-1/m}$, one can check that
the result holds assuming only that
$\ex{w^{-\alpha}}$ is finite for \emph{some} positive $\alpha$.


Finally, we should emphasize that one cannot expect the infinite sum corresponding to
Eq.~(A.7), or the asymptotic expansion of $\ex{\bar{a}_n}$, to converge for any fixed $n$. Indeed, in the log-normal case where $w=e^{-h}$, we have $\mu_1^{(k)}=\ex{w^k}=\exp(\sigma^2(h)k^2/2)$, which becomes $\exp(\sigma^2(h)k(k-1)/2)$ after normalizing so that $\ex{w}=1$. Moreover, the $k$th central moment $\cen{\mu}_1^{(k)}$ also contains a component proportional to $\mu_1^{(k)}$ (it is a linear combination of this and lower moments). Hence, Eq.~(\ref{eq:awxj}) will contain a component proportional to
$\left[\exp\left(\sigma^2(h)/2\right)\right]^{j(j-1)}/n^{j-1}$. For any $n$,
this eventually increases with $j$, so the infinite sum does not converge.
Similar considerations strongly suggest that the first few terms in the asymptotic expansion
will be a good approximation only if $\sigma^2(h)$ is small compared to 
(at most a certain constant times) $\log n$.

\bibliographystyle{unsrt}

\begin{thebibliography}{10}

\bibitem{metr+53jcp}
N.~Metropolis, A.~W. Rosenbluth, M.~N. Rosenbluth, A.~H. Teller, and E.~Teller.
\newblock {Equation of State Calculations by Fast Computing Machines}.
\newblock {\em {J. Chem. Phys.}}, {21}({6}):1087--1092, {1953}.

\bibitem{hast70bio}
W.~K. Hastings.
\newblock {Monte Carlo Sampling Methods Using Markov Chains and Their
  Applications}.
\newblock {\em {Biometrika}}, {57}({1}):--97, {1970}.

\bibitem{frenk-smit02book}
D.~Frenkel and B.~Smit.
\newblock {\em {Understanding Molecular Simulation}}.
\newblock {Academic Press}, {London}, {Second} edition, {2002}.

\bibitem{duan+87plb}
S.~Duane, A.~D. Kennedy, B.~J. Pendleton, and D.~Roweth.
\newblock {Hybrid Monte Carlo}.
\newblock {\em {Phys. Lett. B}}, {195}({2}):216--222, {1987}.

\bibitem{kuma+92jcc}
S.~Kumar, J.~M. Rosenberg, D.~Bouzida, R.~H. Swendsen, and P.~A. Kollman.
\newblock {The weighted histogram analysis method for free-energy calculations
  on biomolecules. I. The method}.
\newblock {\em {J. Comp. Chem.}}, {13}({8}):1011--1021, {1992}.

\bibitem{bono+09jcc}
M.~Bonomi, A.~Barducci, and M.~Parrinello.
\newblock {Reconstructing the equilibrium Boltzmann distribution from
  well-tempered metadynamics}.
\newblock {\em {J. Comp. Chem.}}, {30}({11}):1615--1621, {2009}.

\bibitem{barb+98npb}
I.~M. Barbour, S.~E. Morrison, E.~G. Klepfish, J.~B. Kogut, and M.~P. Lombardo.
\newblock {Results on finite density QCD}.
\newblock {\em {Nucl. Phys. B}}, {60}({1-2}):220--233, {1998}.

\bibitem{fodo-katz02plb}
Z.~Fodor and S.~D. Katz.
\newblock {A New method to study lattice QCD at finite temperature and chemical
  potential}.
\newblock {\em {Phys. Lett. B}}, {534}({1-4}):87--92, {2002}.

\bibitem{smit-robe93jrssb}
A.~F.~M. Smith and G.~O. Roberts.
\newblock {Bayesian computation via the Gibbs sampler and related Markov chain
  Monte Carlo methods}.
\newblock {\em {J. Royal Stat. Soc. B}}, {55}({1}):3--23, {1993}.

\bibitem{feyn-hibb65book}
R.~P. Feynman and A.~R. Hibbs.
\newblock {\em {Quantum Mechanics and Path Integrals}}.
\newblock {McGraw-Hill}, {New York}, {1964}.

\bibitem{parr-rahm84jcp}
M.~Parrinello and A.~Rahman.
\newblock {Study of an F center in molten KCl}.
\newblock {\em {J. Chem. Phys.}}, {80}:860, {1984}.

\bibitem{cepe95rmp}
D.~M. Ceperley.
\newblock {Path integrals in the theory of condensed helium}.
\newblock {\em {Rev. Mod. Phys.}}, {67}({2}):279--355, {Apr} {1995}.

\bibitem{taka-imad84jpsj}
M.~Takahashi and M.~Imada.
\newblock {Monte Carlo calculation of quantum systems. II. Higher order
  correction}.
\newblock {\em {J. Phys. Soc. Jap.}}, {53}({11}):3765--3769, {1984}.

\bibitem{suzu95pla}
M.~Suzuki.
\newblock {Hybrid exponential product formulas for unbounded operators with
  possible applications to Monte Carlo simulations}.
\newblock {\em {Phys. Lett. A}}, {201}({5-6}):425--428, {1995}.

\bibitem{chin97pla}
S.~A. Chin.
\newblock {Symplectic integrators from composite operator factorizations}.
\newblock {\em {Phys. Lett. A}}, {226}({6}):344--348, {1997}.

\bibitem{jang+01jcp}
S.~Jang, S.~Jang, and G.~A. Voth.
\newblock {Applications of higher order composite factorization schemes in
  imaginary time path integral simulations}.
\newblock {\em {J. Chem. Phys.}}, {115}({17}):7832--7842, {2001}.

\bibitem{yama05jcp}
T.~M. Yamamoto.
\newblock {Path-integral virial estimator based on the scaling of fluctuation
  coordinates: Application to quantum clusters with fourth-order propagators}.
\newblock {\em {J. Chem. Phys.}}, {123}:104101, {2005}.

\bibitem{schu96book}
L.~S. Schulman.
\newblock {\em Techniques and applications of path integration}.
\newblock John Wiley, 1996.

\bibitem{chan-woly81jcp}
D.~Chandler and P.~G. Wolynes.
\newblock {Exploiting the isomorphism between quantum theory and classical
  statistical mechanics of polyatomic fluids}.
\newblock {\em {J. Chem. Phys.}}, {74}({7}):4078--4095, {1981}.

\bibitem{chin06pla}
S.~A. Chin.
\newblock {A fundamental theorem on the structure of symplectic integrators}.
\newblock {\em {Phys. Lett. A}}, {354}({5-6}):373--376, {2006}.

\bibitem{li-brou87jcp}
X.~P. Li and J.~Q. Broughton.
\newblock {High-order correction to the Trotter expansion for use in computer
  simulation}.
\newblock {\em {J. Chem. Phys.}}, {86}:5094, {1987}.

\bibitem{ceri+11jcp}
M.~Ceriotti, D.~E. Manolopoulos, and M.~Parrinello.
\newblock {Accelerating the convergence of path integral dynamics with a
  generalized Langevin equation}.
\newblock {\em {J. Chem. Phys.}}, {134}({8}):084104, {2011}.

\bibitem{habe+09jcp}
S.~Habershon, T.~E. Markland, and D.~E Manolopoulos.
\newblock {Competing quantum effects in the dynamics of a flexible water
  model.}
\newblock {\em {J. Chem. Phys.}}, {131}({2}):024501, {2009}.

\bibitem{mark-mano08cpl}
T.~E. Markland and D.~E. Manolopoulos.
\newblock {A refined ring polymer contraction scheme for systems with
  electrostatic interactions}.
\newblock {\em {Chem. Phys. Lett.}}, {464}({4-6}):256, {2008}.

\bibitem{ceri+10jcp}
M.~Ceriotti, M.~Parrinello, T.~E. Markland, and D.~E. Manolopoulos.
\newblock {Efficient stochastic thermostatting of path integral molecular
  dynamics}.
\newblock {\em {J. Chem. Phys.}}, {133}({12}):124104, {2010}.

\bibitem{buss+07jcp}
G.~Bussi, D.~Donadio, and M.~Parrinello.
\newblock {Canonical sampling through velocity rescaling}.
\newblock {\em {J. Chem. Phys.}}, {126}({1}):014101, {2007}.

\bibitem{pred+03jcp}
C.~Predescu, D.~Sabo, J.~D. Doll, and D.~L. Freeman.
\newblock {Heat capacity estimators for random series path-integral methods by
  finite difference schemes}.
\newblock {\em J. Chem. Phys.}, 119(23):12119--12128, 2003.

\bibitem{fent60iret}
L.~Fenton.
\newblock {The sum of log-normal probability distributions in scatter
  transmission systems}.
\newblock {\em {IRE Trans. Comm. Sys.}}, {8}({1}):57--67, {1960}.

\end{thebibliography}

\end{document}